\documentclass[twocolumn]{aastex631}

\shorttitle{CRTS J093417-174421}
\shortauthors{R. Baptista et al.}

\begin{document}

\title{The eclipsing novalike cataclysmic variable 
CRTS~SSS100505~J093417-174421}
 
\author[0000-0001-5755-7000]{Raymundo Baptista}
\affiliation{Departamento de F\'{\i}sica,
Universidade Federal de Santa Catarina,
Campus Trindade,
82912-600 Florian\'opolis - SC, Brazil}
\email{raybap@gmail.com}

\author{Albert Bruch}
\affiliation{Laborat\'orio Nacional de Astrof\'{\i}sica,
Rua Estados Unidos, 154,
37504-364 Itajub\'a - MG, Brazil}
\email{albert@lna.br}
\author[0000-0002-6211-7226]{Raimundo Lopes de Oliveira}
\affiliation{Departamento de F\'{\i}sica,
Universidade Federal de Sergipe, 
Av.\ Marechal Rondon S/N, 
49100-000 S\~ao Crist\'ov\~ao - SE, Brazil}
\affiliation{Observat\'orio Nacional, Rua Gal.\ Jos\'e Cristino, 77,
20921-400 Rio~de~Janeiro - RJ, Brazil}
\affiliation{Departamento de Astronomia, Instituto de Astronomia, 
Geof\'{\i}sica e Ci\^encias Atmosf\'ericas da USP, Cidade Universit\'aria, 
05508-900 S\~ao Paulo, SP, Brazil}
\email{raimundo.lopes@academico.ufs.br}

\author[0000-0002-9459-043X]{Cl\'audia V.\ Rodrigues}
\affiliation{Divis\~ao de Astrof\'{\i}sica, 
Instituto Nacional de Pesquisas Espacias, 
12227-010 S\~ao Jos\'e dos Campos - SP, Brazil}
\email{claudia.rodrigues@inpe.br}

\author[0000-0001-6422-9486]{Alexandre S.\ Oliveira}
\affiliation{IP\&D, Universidade do Vale do Para\'{\i}ba,
12244-000 S\~ao Jos\'e dos Campos - SP, Brazil}
\email{alexandre@univap.br}

\author[0000-0001-6013-1772]{Isabel J. Lima}
\affiliation{Universidade Estadual Paulista,
Campus de Guaratinguet\'a,
Av. Dr. Ariberto Pereira da Cunha, 333
12516-410 Guaratinguet\'a - SP, Brazil}
\email{isabellima01@gmail.com}

\begin{abstract}

Time-resolved optical photometry, complemented by TESS data and long-term
survey light curves, reveals that the transient object
CRTS~SSS100505~J093417-174421 is an eclipsing novalike cataclysmic 
variable of the VY~Scl subtype, with an orbital period of 0.16329188(8)~d.
An analysis of the light curves with eclipse mapping techniques and
an entropy landscape procedure
indicates an orbital inclination of $81.5\degr$ and a mass ratio of 0.45.
Eclipse maps reveal two diametrically opposed
asymmetric arcs of enhanced emission in the intermediate and outer regions
of an accretion disk elongated in the direction perpendicular to the line
joining both stars, interpreted as tidally-induced spiral shock arms. The
accretion disk is 50 per cent larger in the longer wavelength TESS data
than in the optical range, in line with the expected radial temperature
gradient of an opaque steady-state disk.
The combination of a small optical disk radius
(of 21 per cent of the orbital separation) and high orbital
inclination explains the relatively faint absolute magnitude of
$M_g = 7.44$ for a novalike variable.

\end{abstract}


\section{Introduction} 

The foundational model of cataclysmic variables (CVs), developed during
the 1960s and 1970s and summarized in the excellent early reviews by 
\citet{Warner76} and \citet{Robinson76}, has been robustly validated and 
continually refined over the following decades. The accumulated knowledge
about CVs $\sim$20 years later is assessed in \citet{Warner95} comprehensive 
monograph. But then the field has grown so vast that subsequent reviews
are all limited to specific topics.
CVs are interacting binary systems, where a late-type star (the secondary)
fills its Roche lobe and transfers mass to a more massive and evolved
companion, a white dwarf (WD). The dynamics of the mass transfer are
primarily determined by the mass transfer rate and the magnetic field
strength of the WD. The interaction between the infalling material and
the magnetic field of the WD is the main mechanism that determines
whether an accretion disk -- either fully developed or disrupted in
its central regions -- can form or not. 

Over the last half century, the number of known or suspected CVs
has increased dramatically, from a few dozen to several thousand. 
This growth has been driven by
large-scale survey projects such as the Sloan Digital Sky Survey 
\citep[SDSS;][]{York00}, Catalina Real-Time Transient Survey 
\citep[CRTS;][]{Drake09}, All-Sky Automated Survey for Supernovae 
\citep[ASAS-SN;][]{Kochanek17}, and Zwicky Transient Facility 
\citep[ZTF;][]{Bellm19, Masci19}. Many of those objects have never
been studied in greater detail, such that often not more than a
fragmentary light curve or a single spectrum are available. A deeper
study of a larger number of objects is crucial for addressing the many
open questions in the field. For example, the evolutionary pathways
and the connection between physical properties
\citep[e.g.,][]{Zorotovic20, Schaefer24}, or the origin and evolution
of magnetic fields in WDs \citep[e.g.,][]{Briggs18, Schreiber21},
remain major unresolved issues.

Another important question concerns the detailed accretion flow and the 
structure of the accretion disk in CVs. Here, indirect imaging techniques 
provide valuable information. Based on the time dependent structure and
radial velocity variations of spectral lines, Doppler tomography
\citep{Marsh01} permits the reconstruction of the distribution of line
emission in a CV. In eclipsing systems, a detailed analysis of the eclipse
profile allows an even more detailed view. The Eclipse Mapping Method,
first introduced by \citet{Horne85}, transforms the shape of the eclipse
light curve into a map of the surface brightness distribution of the
occulted regions. Past applications of this technique to the 
investigation of the structure, the spectrum, and the time evolution
of accretion disks in CVs have enriched our understanding of these
structures with a wealth of details. These include moving heating/cooling
waves during outbursts in dwarf novae, tidally-induced spiral shocks of
emitting gas with sub-Keplerian velocities, elliptical precessing disks
associated with superhumps, and measurements of the radial run of the
disk viscosity through the mapping of the disk flickering sources. For
reviews about this technique and its previous applications and results,
see \citet{Baptista01,Baptista16a}.

One of the systems that has so far remained largely unstudied is
CRTS~SSS100505~J093417-174421 (hereafter CRTS~J0934), identified as a
transient by the CRTS. Based on its long-term CRTS light curve,
\citet{Oliveira20} included the object in their spectroscopic search for 
magnetic CVs. Its spectrum exhibits the typical features of a CV, namely
strong hydrogen Balmer emission lines along with weaker helium emission lines.
If the broad (FWHM = 2300~km/s) feature around 4960~\AA\ in their spectrum
does indeed include a significant contribution from He~II 4686~\AA, its
equivalent ratio to H$\beta$ is nevertheless too low to satisfy the empirical
criteria \citep{Silber92} for a classification as a magnetic CV. The initial
assumption is that CRTS~J0934 is similar to many other non-magnetic CVs, 
probably of the novalike subtype, since the long-term light curve does not 
reveal dwarf-nova outbursts (see Sect.~\ref{Light curves and eclipse
ephemeris}). 

Here, we present a photometric follow-up of CRTS~J0934 that
reveals the system to be eclipsing. We also carry out a detailed analysis
of its accretion disk using maximum-entropy eclipse mapping techniques.

\section{Observations}
\label{Observations}

We observed CRTS~J0934 during 16 nights between 2021, April 9 and June 18
at the 1.6~m Perkin Elmer (PE) and the 0.6~m Boller \& Chivens (BC)
telescopes of the {\it Observat\'orio do Pico dos Dias} (OPD), Brazil.
A summary is given in Table~\ref{Table: obs-log}. To characterize light
variations on time scales between hours and seconds, we aimed to maximize
the total duration of the light curves and the time resolution while
maintaining a signal-to-noise ratio sufficient to preserve details of
variations. Therefore, most observations were taken in white light, which
leads to an effective wavelength roughly similar to the Johnson V-band
\citep{Bruch18}. Integration times between 5 and 10~s at the larger
telescope and of 60~s at the smaller one were chosen. Together with a
read-out time of 2~s this resulted in the time resolution cited in the
table.

Data reduction followed the usual standard procedures. Simple aperture 
photometry was performed relative to the primary comparison star 
UCAC4~362-054814 \citep{Zacharias13}, the constancy of which was verified 
using 4 other comparison stars in the field. The average magnitudes 
quoted in Table~\ref{Table: obs-log} are rough estimates, calculated
by adding the V magnitude of the comparison star as quoted in the UCAC4
catalogue to the mean of the nightly differential white light magnitude
between the target and the comparison star (excluding eclipses). 
Magnitudes were thus not reduced to a standard photometric system, and
their absolute values are not of high precision. However, magnitude
differences between different nights should be more reliable. All time
stamps were transformed into Barycentric Julian Date on the Barycentric
Dynamical Time scale, using the on-line tool of \citet{Eastman10}.

\begin{table}
	\centering
	\caption{Journal of observations at OPD}
\label{Table: obs-log}

\begin{tabular}{lcccccc}
\hline

Date  & Start & $\Delta T^a$ & N$^b$  & $\Delta t^c$ & av.  & Tel.$^d$ \\
2021  & (UT)  & (min)      &    & (sec)      & mag. &          \\
\hline
Apr  09 & 00:34 & 138 &  673 & 12 & 17.97 & PE \\
Apr  10 & 00:19 & 166 &  714 & 12 & 18.13 & PE \\
Apr  11 & 00:03 & 159 & 1263 &  7 & 18.08 & PE \\
Apr  11 & 23:59 & 174 & 1439 &  8 & 18.13 & PE \\
Apr  19 & 22:52 & 203 &  189 & 62 & 18.02 & BC \\
Apr  20 & 22:07 & 288 &  220 & 62 & 18.06 & BC \\
Apr  21 & 21:35 & 318 &  306 & 62 & 18.19 & BC \\
Apr  22 & 21:37 & 320 &  305 & 62 & 18.28 & BC \\
Apr  28 & 21:38 & 308 &  300 & 62 & 18.33 & BC \\
Apr  29 & 21:39 & 249 &  201 & 62 & 18.27 & BC \\
Apr  30 & 21:25 & 256 &  206 & 62 & 18.44 & BC \\
May  01 & 21:34 & 131 &  118 & 62 & 18.18 & BC \\
May  02 & 21:33 & 277 &  270 & 62 & 18.25 & BC \\
May  18 & 21:51 & 201 &  173 & 62 & 18.20 & BC \\
Jun  17 & 21:27 & 102 &  100 & 62 & 18.24 & BC \\
Jun  18 & 21:14 &  88 &   87 & 62 & 18.16 & BC \\ [1ex]
\hline
\multicolumn{7}{l}{$^a$: total time base;} \\
\multicolumn{7}{l}{$^b$: number of integrations;} \\
\multicolumn{7}{l}{$^c$: time resolution;} \\
\multicolumn{7}{l}{$^d$: PE = 1.6m Perkin Elmer; BC = 0.6m Boller \& Chivens} \\
\end{tabular}
\end{table}

CRTS~J0934 was also observed by the
Transiting Exoplanet Survey Satellite \citep[TESS;][]{Ricker14} 
and data are available in the mission's public archive. TESS observes 
sectors of the sky almost continuously during two spacecraft orbits,
spanning about 27 days, only interrupted during short intervals for
data download to Earth. The photometry is conducted in a broad passband, 
from 600 to 1000~nm, centered on the Cousins $I$ band. The public TESS 
data of CRTS~J0934 were downloaded from the Mikulski Archive for Space 
Telescopes (MAST) 
portal.\footnote{https://mast.scsci.edu/portal/Mashup/Clients/Mast/Portal.html}
The system was observed in four sectors (see Table \ref{Table: TESS data}). 
Light curves from sectors 8 and 35 were extracted using the 
\textsc{Lightkurve}\footnote{https://docs.lightkurve.org/index.html}
package. Those of sectors 62 and 89 are available at the MAST site as 
pre-reduced light curves. Simple aperture photometry (SAP) data were used.
Details are listed in Table~\ref{Table: TESS data}. CRTS~J0934 does not
have neighbors closer than $\approx$$40''$. Therefore, even considering the 
large projected pixel size of TESS, contamination of the light curves is 
expected to remain small. 
In any case, its effect on the following analysis is negligible, as the
TESS data were used to determine eclipse epochs, width, and shape, which
are unaffected by additive contamination on timescales longer than the
orbital period.

\begin{table}
	\centering
	\caption{Journal of TESS observations}
\label{Table: TESS data}

\begin{tabular}{ccccc}
\hline
TESS    & Start & End    & Number of         & $\Delta t$ \\
sector  & (BJD) & (BJD)  & exposures & (min)      \\
\hline
08 & 2458517.37 & 2458541.99 &   894 & 30 \\
35 & 2459255.00 & 2459279.98 &  2815 & 10 \\
62 & 2459988.45 & 2460014.16 & 18047 &  2 \\
89 & 2460718.14 & 2460746.85 & 19730 &  2 \\
\hline
\end{tabular}
\end{table}

To investigate the long-term behavior of CRTS~J0934 over approximately 15 
years, we also retrieved its light curves from the CRTS and ZTF. They cover 
the period from 2005 to 2015 (CRTS), and from 2019 to 2024 (ZTF).

\section{Light curves and eclipse ephemeris}
\label{Light curves and eclipse ephemeris}

\begin{figure*}
\figurenum{1}
\label{lightcurves}
\plotone{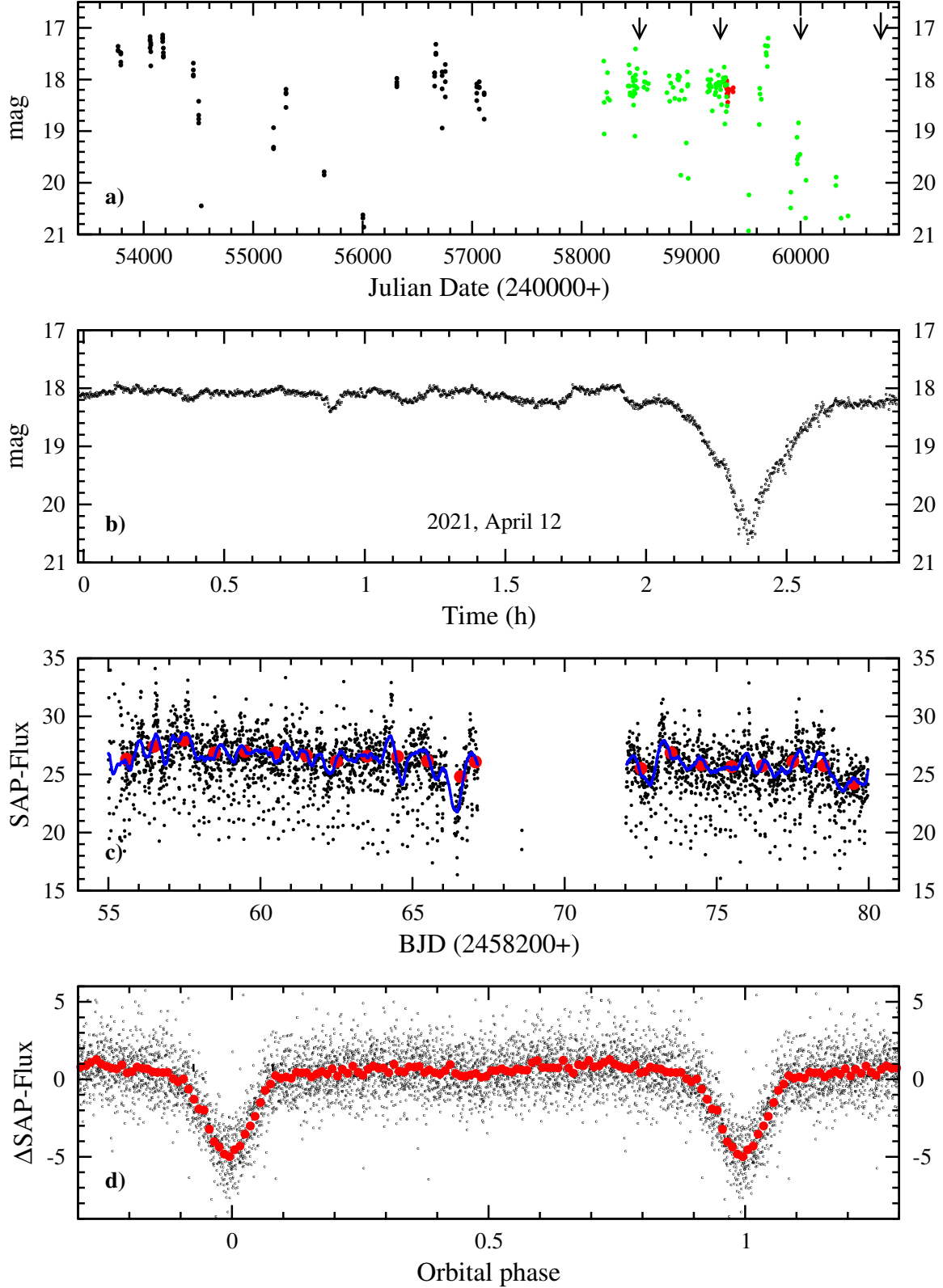}
\caption{{\it a)} Long-term light curve of CRTS~0934. Black and green dots
are CRTS and ZTF data, respectively. The red dots represent the nightly
average out-of-eclipse magnitude of the OPD observations. The black
arrows indicate mid-epochs of the TESS data. {\it b)} Time-resolved OPD
light curve of 2021, April 12 (starting just before midnight of April 11).
{\it c)} Sector 35 TESS light curve. The red dots are average flux values
(after masking eclipses) within 1~d intervals. The light curves smoothed
with a Savitzky-Golay filter (again, eclipses masked) are shown in blue.
{\it d)} Combined TESS light curve of sectors 8 and 35, folded on the 
orbital period (small black dots). The red dots represent average values
in phase bins of width 0.01.}
\end{figure*}

The long-term light curve of CRTS~J0934 is shown in Fig.~\ref{lightcurves}a.
The black and green dots are CRTS and ZTF $g$-band data, respectively. The 
red dots represent average nightly (out-of-eclipse) magnitudes of the OPD 
observations. The black arrows indicate mid-epochs of the TESS data. No 
attempt has been made to adjust the different photometric systems because 
this is not relevant to the purposes of this work. CRTS~J0934 typically 
remained at a magnitude slightly fainter than 18~mag, occasionally 
brightening by about half a magnitude, but sometimes exhibiting low states 
as faint as nearly 21~mag (noting that only a few of the faint data points 
in the figure correspond to eclipse phases). No dwarf-nova outbursts were 
observed. Such behavior is consistent with novalike CVs of the VY~Scl
subtype \citep[e.g.,][]{Leach99, Duffy24}. 

On time scales of minutes (see Fig.~\ref{lightcurves}b for an example), 
CRTS~J0934 is characterized by strong flickering with an amplitude on
the order of 0.3~mag, typical for VY~Scl stars \citep{Bruch21}.
On longer timescales, however, eclipses are the dominant feature, with
10 events recorded in the OPD light curves.
Their average profile is shown in Fig.~\ref{figure2}a
as green dots with error bars.
The eclipse epochs were determined by the minimum of a fit of a fifth-order 
polynomial to the profiles, allowing us to derive a preliminary orbital 
period. As an example of the photometric behavior of CRTS~J0934 over
several days, its sector 35 TESS light curve is displayed in
Fig.~\ref{lightcurves}c. Here, the red dots are average values in 1~d
intervals after masking eclipses. Irregular variations on daily time
scales occur, as is common in novalike CVs 
\citep[see, e.g., Appendix A of][]{Bruch24a}. The TESS light curves are 
too noisy to measure individual eclipse epochs. Even more than in the 
earlier sectors, this is the case for sectors 62 and 89. CRTS~J0934 was 
observed in a low state or in the transition into a low state 
(Fig.~\ref{lightcurves}a) in Sector 62. The high noise level in sector 89 
suggests that the system did not yet recover from the low state after the 
last ZTF data point. Although individual eclipse epochs cannot be measured 
in the TESS light curves, folding the data on the preliminary period
yields an average orbital light curve with clearly visible eclipses, as
is shown in Fig.~\ref{lightcurves}d for the combined sectors 8 and 35 data. 
Before phase-folding, a smoothed version of the light curves (shown in
blue for the sector 35 light curve in Fig.~\ref{lightcurves}c) was
subtracted from the original data to remove variations on supraorbital
time scales. This version was obtained using a Savitzky-Golay 
\citep{Savitzky64} filter with a cutoff time scale of 1~d and a 4$^{\rm th}$ 
order smoothing polynomial. The folded light curves enable the determination
of eclipse epochs for each sector with no cycle count ambiguities, extending
the total time base greatly. They are listed in Table~\ref{Table: ecl-time}. 
All eclipse epochs together then yield the eclipse ephemeris
\begin{equation}
  T_{\rm min} = BJD\, 2459316.6035(2) + 0.16329188(8) \times E,
  \label{eq:ephemeris}
\end{equation}
where $E$ is the cycle number.

\begin{table}
	\centering
	\caption{Eclipse epochs}
\label{Table: ecl-time}

\begin{tabular}{lcc}
\hline
Ecl. number & Ecl. epoch    & Telescope \\
            & BJD (2400000+)&  \\
\hline
$-$5000           & 58500.1449 & TESS \\
\phantom{0}$-$714 & 59200.0131 & TESS \\
\phantom{$-$000}0 & 59316.6028 & PE   \\
\phantom{$-$00}55 & 59325.5849 & BC   \\
\phantom{$-$00}61 & 50326.5646 & BC   \\
\phantom{$-$00}67 & 59327.5429 & BC   \\
\phantom{$-$0}103 & 59333.4227 & PE   \\
\phantom{$-$0}104 & 59333.5867 & BC   \\
\phantom{$-$0}110 & 59334.5665 & BC   \\
\phantom{$-$0}116 & 59335.5462 & BC   \\
\phantom{$-$0}128 & 59337.5044 & BC   \\
\phantom{$-$0}226 & 59353.5065 & BC   \\
\phantom{$-$}4112 & 59988.0582 & TESS \\
\phantom{$-$}8383 & 60718.1389 & TESS \\
\hline
\end{tabular}
\end{table}

\section{Eclipse tomography}
\label{Eclipse tomography}

We apply maximum-entropy eclipse mapping techniques to average
OPD and TESS eclipse light curves of CRTS~J0934. This
analysis allows us to derive the surface brightness distribution of
its accretion disk at two different wavelengths and the 
uneclipsed component of the light curves (light from the secondary
star and/or a vertically-extended disk wind), as well as to identify
and investigate structures within the disk.

As a first step, we obtained average OPD and TESS light curves.
The OPD data cover a relatively short time span in which CRT~J9034 was
in its typical brightness level and consistently showed the same eclipse
shape. To improve the signal-to-noise and to minimize the influence of
flickering for the subsequent analysis, the individual OPD light curves
were phase-folded according to the ephemeris of Eq.\,(\ref{eq:ephemeris})
and combined in an average eclipse light curve with a resolution of 0.005
in phase. No data segment was removed and no filtering was applied to the
OPD data before combining the light curves. A median flux was computed
for each phase bin; the median of the absolute deviations with respect
to the median was taken as the corresponding uncertainty at each bin.
The resulting average light curve has signal-to-noise ratios ranging
from S/N $\simeq$ 40 (out-of-eclipse) to $\simeq$ 15 (mid-eclipse).
A similar procedure was applied to the combined TESS
light curve of the sectors 8 and 35. In this case, we compensated
for the subtraction of a smoothed version of the light
curves before phase-folding (see Fig.~\ref{lightcurves}c)
by adding an arbitrary offset to the data in order to ensure positive
fluxes across the
eclipse.\footnote{The logarithmic definition of entropy in the eclipse
  mapping method \citep{Horne85} requires intensities in the eclipse map
  and fluxes in the light curves to always be positive numbers.}
The data were further renormalized to a unity flux eclipse depth before
computing the median flux at each phase bin. The resulting average TESS
light curve has lower signal-to-noise, ranging from S/N= 20
(out-of-eclipse) to $\simeq 7$ (mid-eclipse). Because of the arbitrary
offset added to the average TESS light curve, the uneclipsed component
in this case is meaningless. Hereafter, we will refer to the average
OPD and TESS light curves, respectively, as the V-band and I-band light
curves to emphasize their different wavelengths.

Second, the eclipse geometry for the image reconstructions needs to
be selected. The eclipse geometry is defined by a pair of values of the
orbital inclination $i$ and the mass ratio $q=M_2/M_1$ (where $M_1$ is
 the WD mass and $M_2$ is the secondary star mass) or, alternatively, 
the duration $\Delta\phi$ of the eclipse of the disk center. For a
fixed $\Delta\phi$ value, there is a unique relation between $i$ and
$q$ \citep{Bailey79, Horne85}. In low-$\dot{M}$ systems, such as
quiescent dwarf novae, $\Delta\phi$ can be directly measured from the
light curve by the ingress/egress phases of the WD eclipse. However,
in novalike systems such as CRTS~J0934, the accretion disk is much
brighter than the WD and veils its contribution to the eclipse shape.
In these cases, $\Delta\phi$ can be estimated from the eclipse full
width-at-half depth (FWHD), or the full width-at-half maximum flux
level (FWHM). This is based on the notion that the shadow of the
secondary star covers about half of the entire disk when the disk
center is just eclipsed or just reappeared from eclipse (FWHM) or,
alternatively, that at these phases it covers about half of all the
light that is hidden during eclipse minimum (FWHD). From the
higher S/N, average V-band light curve of CRTS~J0934,
we estimate $\mathrm{FWHD}= 0.078$ and $\mathrm{FWHM}=0.087$ (in units
of the orbital period).

\begin{figure}
\figurenum{2}
\label{figland}
\plotone{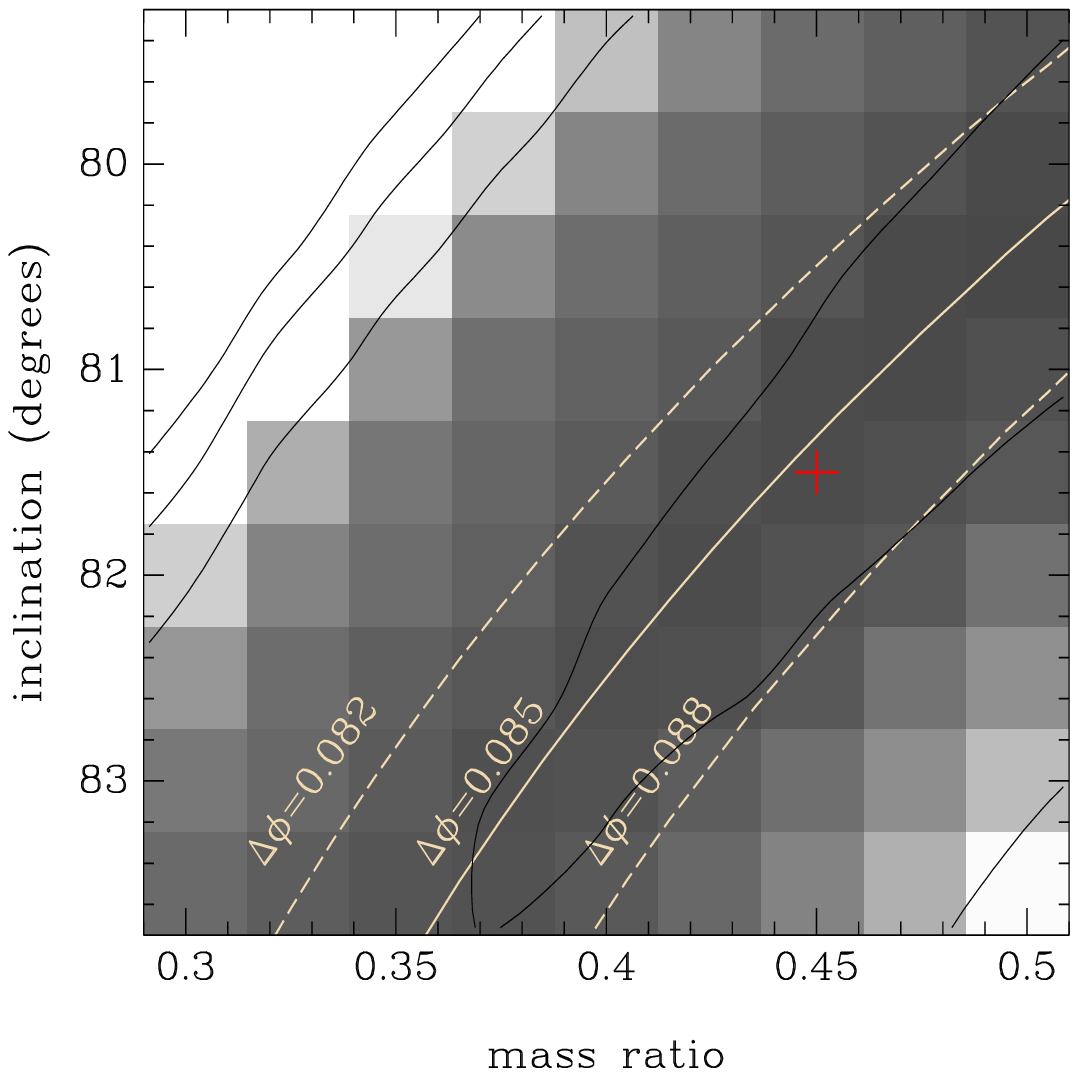}
\caption{Entropy landscape in the ($i,q$) binary parameter plane.
  The entropy of the eclipse maps are shown as different shades of gray;
  darker regions have higher entropy. Contour lines indicate the shape of
  the entropy landscape and a red cross marks the location of highest
  entropy. White curves show lines of constant $\Delta\phi$ values.}
\end{figure}

This estimate can be further improved with an ``entropy landscape''
procedure \citep[e.g., ][]{Baptista16}. For a given $\Delta\phi$ value,
the surface brightness distribution is skewed towards (away from) the
L1 point if the inclination is underestimated (overestimated). Therefore,
an incorrect inclination increases the amount of structure and the
deviations from axi-symmetry of the eclipse map, which is flagged with
a lower entropy value. Hence, the best choice for the eclipse geometry
can be inferred by computing eclipse maps for a range of $(i,q)$ values
and finding the pair of values that yields the eclipse map of highest
entropy. We again used the V-band light curve for this procedure.
The results of the entropy landscape are shown in Fig.~\ref{figland}. The
location of highest entropy is marked by a red cross. The maximum-entropy
geometry parameters are $i= 81.5\degr \pm 1.0\degr$ and $q=0.45 \pm 0.05$,
corresponding to an eclipse width of $\Delta\phi= 0.085 \pm 0.003$ in
phase, consistent with the range for these parameters inferred from the
eclipse FWHD and FWHM estimates. This $(i,q)$ pair of values will be
adopted hereafter.

\subsection{The CRTS~J0934 eclipse maps}
\label{The J0934 eclipse map}

We use an eclipse map consisting of a flat Cartesian grid in the orbital
plane centered on the position of the WD, with $51\times 51$ pixels and
side $2\,R_\mathrm{L1}$ (where $R_\mathrm{L1}$ is the distance from the
disk center to the inner Lagrangian point L1). Our eclipse mapping code
implements a scheme of double default functions, $D_{+} D_{-}$, by
simultaneously steering the solution towards the most nearly axi-symmetric
map consistent with the data ($D_{+}$) and away from the crisscrossed arcs
along the edges of the shadow of the occulting secondary star ($D_{-}$)
\citep{Spruit94,Baptista05}. It is optimized to recover asymmetric
structures such as spirals arms \citep[see, e.g.,][]{Harlaftis04}. The
positive default function is a polar Gaussian with radial and azimuthal
blur widths respectively of $\Delta r= 0.02\,R_\mathrm{L1}$ and
$\Delta\theta= 20\degr$ in order to minimize the azimuthal
smearing effect and to improve the reconstruction of asymmetric
sources.\footnote{This choice for $\Delta\theta$ implies that any
  off-center point source is smeared in azimuth by $\simeq 60\degr$.}
The negative default function is a Gaussian along the ingress/egress arcs
of phase width $d\phi= 0.01$.

\begin{figure*}
\figurenum{3}
\label{figure2}
\includegraphics[width=0.76\textwidth,angle=270]{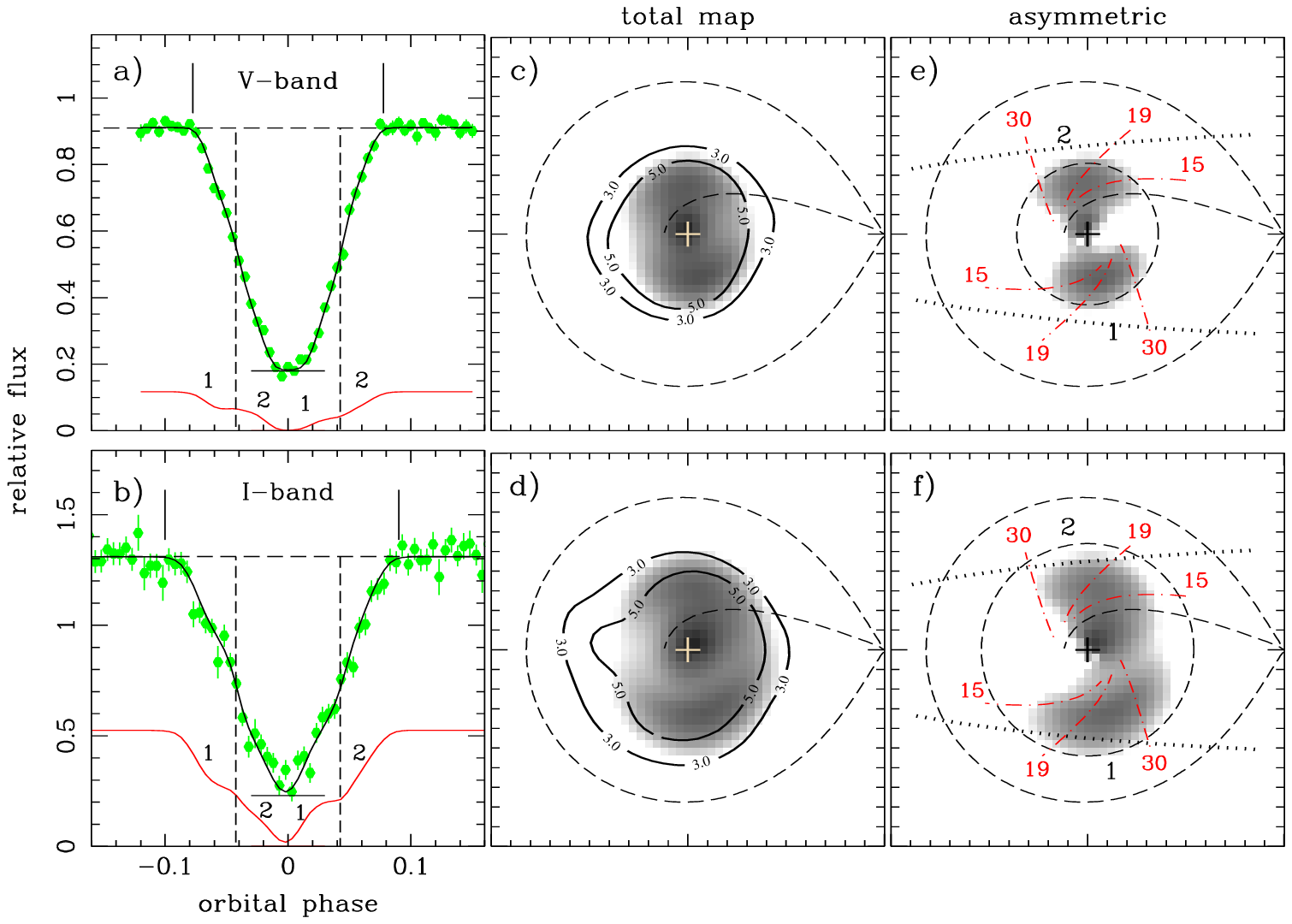}
\caption{Left-hand panels: average data (green points with error bars)
  and best-fit eclipse mapping model (solid line) 
  for the V-band (a) and I-band (b) light curves.
  Horizontal bars at mid-eclipse show the uneclipsed flux {\bf in
  each case}. Vertical dashed lines mark the best-fit $\Delta\phi$ value
  (indicating the WD ingress/egress phases), and vertical
  ticks mark the total width of the eclipse, $\Delta\phi_E$. 
  The lower red curve illustrates the eclipse light curve of the 
  asymmetric arcs 1 and 2 (see text).
  Middle panels: V-band (c) and I-band (d) eclipse maps
  in a logarithmic grayscale; darker regions are brighter.
  Regions inside the two solid contour lines are above the
  3$\sigma$ and {\bf 5$\sigma$} confidence levels, respectively.
  A cross marks the position of the disk center; dashed lines depict the
  primary Roche lobe and the gas stream trajectory.  
  Right-hand panels: The asymmetric component of the
  V-band (e) and I-band (f) eclipse maps in a logarithmic grayscale.
  Dashed circles mark the estimated disk radii; dotted lines show the
  edge of the projected shadow of the secondary star
  at phase zero, while dot-dashed red lines illustrate the expected
  orientation of tidally-induced spiral arms for constant opening
  angles of $\theta_s$ = 15$\degr$, 19$\degr$, and 30$\degr$.
  The location of asymmetric arcs 1 and 2 is indicated.}
\end{figure*}

Because this version of the eclipse mapping method does not take into
account for out-of-eclipse brightness changes, these were removed by
dividing the average light curve by the best-fit spline function to the
out-of-eclipse portion of the light curve, and scaling the result to the
average out-of-eclipse flux level. With the assumption that the eclipse
mapping model provides a complete description of the data, the error bars
were further uniformly scaled to ensure a reduced $\chi^2 = 1$ for the
reconstructions.\footnote{This also ensures that the Monte Carlo
  procedure properly evaluates the uncertainties in the eclipse map.}
The resulting light curves are shown in the left-hand panels
of Fig.~\ref{figure2} as green dots. 
They show the characteristics double-stepped shape seen in
 the light curves of IP~Peg in outburst, indicating the progressive
 occultation of two brightness sources, none of which coincide with disk
 center.

The uncertainties in the eclipse map were derived from Monte Carlo
simulations with the average eclipse light curve, generating a set of 20
randomized eclipse maps \citep[see][]{Rutten92}. These are combined to
produce a map of the standard deviations with respect to the true map.
A map of the statistical significance (or the inverse of the relative error)
is obtained by dividing the true eclipse map by the map of the standard
deviations \citep{Baptista05}. The uncertainties obtained with this
procedure are used to plot the confidence level contours of the eclipse
maps in the middle panels of Fig.~\ref{figure2} and to
estimate the uncertainties of the azimuthal distributions of
Fig.~\ref{spiral}.

The solid lines in the left-hand panels of Fig.~\ref{figure2}
are the eclipse light curves resulting from the best fit eclipse maps
depicted on a logarithmic grayscale in the middle panels of
Fig.~\ref{figure2}. The orientation of the eclipse map is such that the
secondary star rotates counter-clockwise around the WD at the center
of the map (in the observer's reference frame), while the observer 
rotates clockwise (in the binary reference frame). At phase zero, the
secondary and the observer are on the right side of the map (with the
projected shadow of the secondary star completely covering the disk).
At phase $+0.25$ the observer is at the bottom (in the binary frame)
while the secondary star is at the top of the eclipse map (in the
observer's frame).

The eclipse maps in the middle panels of Fig.~\ref{figure2}
show no evidence of a bright spot at the disk rim, and no enhanced
emission along the gas stream trajectory. The statistical significance of
the surface brightness distributions is above the 5$\sigma$
confidence level. In the V-band,
the accretion disk is small and fills
less than half of the primary Roche lobe. From the measured total eclipse
duration of $\Delta \phi_E = 0.156 \pm 0.001$ in phase, its radius is
estimated to be $R_d/R_{\rm L1} = 0.36$.
The I-band eclipse is wider, with $\Delta\phi_E= 0.200 \pm
0.005$, which leads to a larger estimated radius of $R_d/R_{L1}= 0.54$,
50 per cent larger than in the V-band. This is in line with the expected
behavior of an opaque steady-state disk, in which the temperature drops
with distance from disk center as $T(R)\propto R^{-3/4}$.

The right-hand panels of Fig.~\ref{figure2} show
the asymmetric component of the eclipse maps,
obtained by slicing the disk in a set of radial bins and by fitting a
smooth spline function to the median of the lower half of the intensities
in each bin. Then the spline-fitted intensity in each bin is subtracted
from the total intensity. This procedure essentially separates the baseline
of the radial profile from the azimuthal structures into symmetric and
asymmetric maps, respectively. A dashed circle shows the inferred 
disk radius in each case.
Dotted lines depict the shadow of the secondary star at phase zero,
indicating that in the V-band the disk is fully eclipsed
at this phase. Given that all V-band light from the
accretion disk is occulted at phase zero, the remaining flux at that
phase cannot be accommodated in the eclipse map and is dumped into the
uneclipsed flux component. We find a significant uneclipsed component
of 20 per cent of the total out-of-eclipse V-band flux
(indicated by the horizontal bar at mid-eclipse in Fig.~\ref{figure2}a).

\subsection{Spiral structures}
\label{Spiral structures}

Apart from the central brightness enhancement, both in
the V-band and in the I-band the accretion disk is elongated in the
direction perpendicular to the line joining both stars and displays
two asymmetric, arc-shaped sources of comparable brightness, located
in the lower right (arm 1) and upper left (arm 2) quadrants of the
eclipse maps in Fig.~\ref{figure2}. The arcs show an azimuthal extent
of $\sim 90\degr$, extend from intermediate ($R \simeq 0.2\;
R_\mathrm{L1}$) to outer disk regions, and represent a
significant fraction of the emitted light. They account for 33 per cent
of the total out-of-eclipse flux (or 40 per cent of the disk flux) in
the V-band and for 40 per cent of the total out-of-eclipse flux (or 48
per cent of the disk flux) in the I-band.
These features resemble the asymmetric structures observed in the eclipse
maps of IP~Peg \citep{Baptista00, Baptista02, Baptista05}, 
V348~Pup \citep{Saito16} and V2051~Oph \citep{Baptista20}, which have
been interpreted as spiral shocks in the outer disk regions, induced by
tidal forces from the secondary star \citep{Sawada86a, Boffin01, Steeghs01}.
For more details, see Sect.~\ref{Spiral shocks}.

\begin{figure}
\figurenum{4}
\label{spiral}
\includegraphics[width=0.37\textwidth,angle=270]{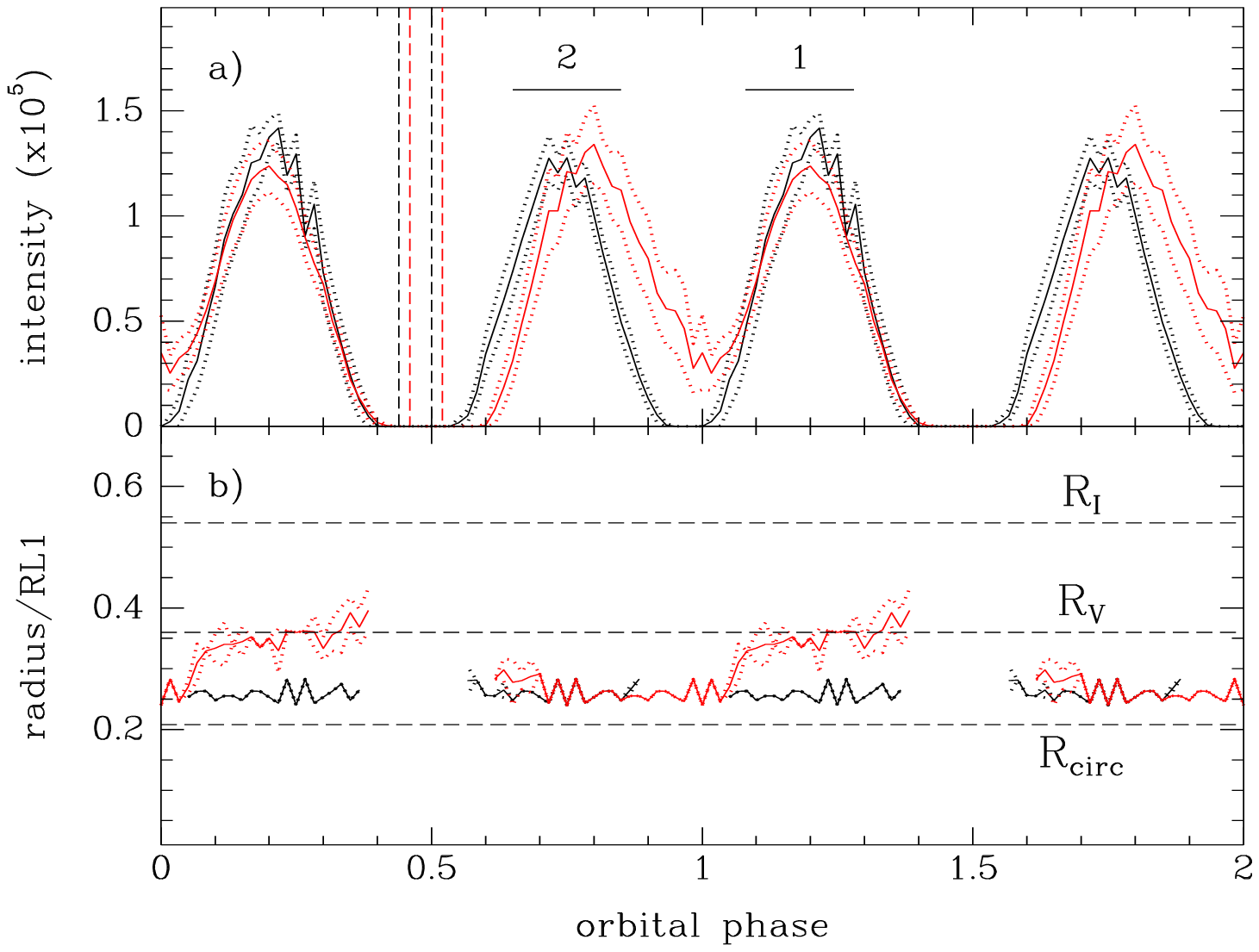}
\caption{{\it a)} Dependency on orbital phase of the maximum intensity
    derived from the asymmetric component of the V-band 
    (black) and I-band (red) eclipse maps. Intensities are plotted 
    on an arbitrary scale; arms 1 and 2 are labeled and horizontal tick 
    marks indicate the phases of maxima of the double-wave orbital
    modulation. Dotted lines show the uncertainties at the 1$\sigma$
    limit. Vertical dashed lines mark the 1-$\sigma$
    range of values for the phase of lowest intensity,
    $\phi_1 (I_\mathrm{min})$ (see text).
    {\it b)} Corresponding radius of maximum intensity in units of
    $R_\mathrm{L1}$. Horizontal dashed lines mark the circularization
    radius ($R_\mathrm{circ}= 0.208\,R_\mathrm{L1}$) and the outer disk
    radius ($R_V= 0.36\,R_\mathrm{L1}$ in the V-band and
    $R_I= 0.54\,R_\mathrm{L1}$ in the I-band). }
\end{figure}

To investigate the properties of the asymmetric arcs, we followed the
method of \cite{Baptista00}. We divided the eclipse map into azimuthal
slices (i.e., ``pizza slices'') and computed the radius at which the 
intensity reaches a maximum for each azimuth. A corresponding Keplerian 
velocity is computed for the radius of maximum intensity, assuming 
$M_1=0.66\,M_\odot$ and $R_\mathrm{L1}= 0.72\,R_\odot$ (see
Sect.~\ref{Discussion}). This exercise allows us to recover the
location of the spiral structures both in radius and in azimuth
\citep[e.g.,][]{Baptista00, Baptista05}. The results are shown in
Fig.~\ref{spiral} as a function of binary phase. The distribution of
the maximum intensity $I_\mathrm{max}(\phi)$ from the azimuthal slices
and the corresponding radius $R(I_\mathrm{max})$, are shown in the
upper and lower panels, respectively.
The uncertainties in these two quantities are obtained from the
 Monte Carlo simulations described in Sect.~\ref{The J0934 eclipse map}.
Orbital phases are measured from the line connecting the two stars and
increase in the clockwise direction (Fig.~\ref{figure2}). The two-armed 
asymmetric structures lead to a double-humped shape in the intensity
distribution, with the maxima indicating the azimuthal positions of
arms 1 and 2.
In the V-band, arms 1 and 2 lay at a comparable distance of
 $R\simeq 0.26\,R_\mathrm{L1}$ from the disk center (expected Keplerian
velocities of about $800-840$~km/s); in the I-band, Arm 1 is farther
away, at $R\simeq 0.35\,R_\mathrm{L1}$ (corresponding to Keplerian
velocities of $\simeq 690-720$~km/s).
These velocities, corresponding to a location not far from the outer disk
edge, should be lower than the velocity determined from the integrated
disk light. They are thus compatible with FWHM of 1130~km/s (1260~km/s) of 
the H$\alpha$ (H$\beta$) emission lines reported by \citet{Oliveira20}.
Dashed lines in the lower panel of Fig.~\ref{spiral} depict the
circularization radius for $q=0.45$ ($R_\mathrm{circ}= 0.208\,
R_\mathrm{L1}$) and the estimated outer disk radius  
in each band. The two-armed structures are within these two limits.
Spiral shocks are expected to exhibit sub-Keplerian velocities, with a
velocity reduction that can be as large as 30 per cent \citep[e.g., ][]
{Baptista00}. If this is the case, the hot, post-shock, line emitting
gas is expected to appear as arc-shaped structures at velocities of
$\simeq 550$~km/s in Doppler tomograms.

\section{Discussion}
\label{Discussion}

\subsection{Basic system parameters}
\label{Basic system parameters}

We have detected eclipses in the nova-like CV CRTS~J0934 through
phase-resolved ground and space-based optical observations spanning more
than six years. These eclipses permit a high-precision measurement of
the system's orbital period. At about 3.92~h, it falls within the
typical range for nova-like variables. 

A detailed analysis of the average eclipse profile using eclipse mapping
techniques reveals further details of the structure of CRTS~J0934. It
turns out that the V-band accretion disk radius of
$R_V/R_{\rm L1} = 0.36$ is relatively small compared to that of other
nova-like CVs. The approximations of \citet{Plavec64} and
\citet{Eggleton83} together with the mass ratio of 0.45 yield a
volume equivalent radius of the WD Roche lobe of 
$R_{\rm 1,eq} = 0.78\, R_{\rm L1}$. The CRTS~J0934 disk radius is then 
$0.46\, R_{\rm 1,eq}$. \citet{Harrop-Allin96} determined the disk radii
in 25 nova-like variables and old novae, finding values ranging between 
$0.46 \le R_d/R_{\rm 1,eq} \le 0.95$. Thus, the V-band
radius of the CRTS~J0934 disk lies at the lower edge of the disk radii
range measured by \citet{Harrop-Allin96}. 

The small V-band disk radius, combined with the high
orbital inclination of $81.5\, \degr$, accounts for the faint 
optical absolute magnitude of $M_g = 7.44 \pm 0.57$
derived for the system. This value was obtained from the average
apparent magnitude of $g = 18.69 \pm 0.28$, assuming a distance
of $1646^{+457}_{-301}$~pc \citep{Bailer-Jones21} and an
interstellar extinction of $A_V = 0.159$, using the 3D extinction
maps of \citet{Doroshenko24} together with a ratio $A_V/A_g = 0.94$
\citep{Schlafly11}. Applying the correction suggested by 
\citet{Paczynski80} to adjust the disk magnitude to the average
inclination (while neglecting any contribution from the secondary star
to the total light) yields an  absolute magnitude of $M_{g,{\rm corr}} =
5.72 \pm 0.57$. Reflecting the effect of the small V-band
disk radius, this value lies at the faint end of the absolute magnitude
distribution of novalike variables above the CV period gap \citep{Bruch24b}.

Some important system parameters have already been derived in the 
preceding sections. Others can be inferred based on reasonable assumptions. 
They are listed in Table~\ref{Table: system parameters}. The fundamental 
assumption concerns the mass of the secondary star, which is taken to be 
$M_2 = 0.297\, M_\odot$, according to the empirical relation between orbital 
period and $M_2$ of \citet{Knigge11}. We adopted a conservative uncertainty 
of 20 per cent to the $M_2$ estimate. Together with the adopted mass ratio, 
this yields a WD mass of $M_1 = 0.66\, M_\odot$. The component separation 
$a$ follows from Kepler's third law. The approximation of \citet{Plavec64} 
for $R_{\rm L1}/a$ then yields the distance between the WD and the L1 point. 
The mass-radius relation for WDs of \citet{Nauenberg72} provides the radius 
of the primary star, while $R_2$ is taken as the volume equivalent Roche 
lobe radius of the secondary star, following \citet{Eggleton83}. The outer 
disk radius is estimated from the measured total eclipse duration 
(Sect. \ref{The J0934 eclipse map}). Standard error propagation and Monte 
Carlo simulations with $10^5$ trials on the input parameters $q$, $M_2$ and 
$\Delta\phi_E$ is used to estimate the uncertainties of the remaining binary 
parameters.

\begin{table}
	\centering
	\caption{System parameters for CRTS~J0934}
\label{Table: system parameters}

\begin{tabular}{llc}
\hline
orbital period         & $P_{\rm orb}$ & $0.16329188 \pm 0.00000008$~d \\
apparent magnitude$^a$ & $m_g$       & $18.69 \pm 0.28$              \\
absolute magnitude$^a$ & $M_g$       & $7.44 \pm 0.57$               \\
distance$^b$           & $d$         & $1645 \pm 379$~pc             \\
orbital inclination    & $i$         & $81.5\degr \pm 1.0\degr$      \\
component separation   & $a$         & $1.238 \pm 0.088\, R_\odot$   \\
distance WD -- L1      & $R_{\rm L1}$ & $0.716 \pm 0.059\, R_\odot$  \\
accretion disk radius  & $R_V$       & $0.36 \pm 0.01\,R_{\rm L1}$   \\
                       & $R_I$       & $0.54 \pm 0.05\,R_{\rm L1}$   \\
mass ratio             & $q$         & $0.45 \pm 0.05$              \\
primary star mass      & $M_1$       & $0.66 \pm 0.15\, M_\odot$    \\
primary star radius    & $R_1$       & $0.012 \pm 0.002\, R_\odot$  \\
secondary star mass    & $M_2$       & $0.297 \pm 0.059\, M_\odot$  \\
secondary star radius  & $R_2$       & $0.386 \pm 0.027\, R_\odot$  \\
rad.\ vel.\ amplitude  & $K_1$       & $118 \pm 10$~km/s          \\
                       & $K_2$       & $262 \pm 24$~km/s          \\
\hline
\multicolumn{3}{l}{$^a$: average out-of-eclipse magnitude in high state}\\
\multicolumn{3}{l}{$^b$: \citet{Bailer-Jones21}}\\
\end{tabular}
\end{table}

Our eclipse mapping revealed that about 20 per cent of the total 
V-band out-of-eclipse flux remains uneclipsed. The source of this
extra light cannot be in the orbital plane. For a secondary star mass
of $0.297\ M_\odot$, \citet{Knigge11} list an absolute magnitude of 
$M_V = 10.93$. Comparing this to the absolute magnitude of the entire 
system of $M_g = 5.72$, it is obvious that the secondary star can account
for only a small fraction of the uneclipsed light. This 
could be different if the secondary were evolved. To fit in the Roche-lobe of
a $\sim$4~h binary, it then must have been stripped of its outer layers, 
exposing its hotter interior. Depending on the degree of stripping, nuclear 
processed matter may then be transferred to the primary component. A detailed 
spectroscopic analysis may confirm or discard this hypothesis. Alternatively,
most of the uneclipsed light may have a different origin, such as, for 
instance, a vertically extended disk wind.

\subsection{Spiral shocks}
\label{Spiral shocks}

Another noteworthy feature is the presence of two asymmetric arcs in the
accretion disk. Numerical simulations by \cite{Meglicki93} reveals that
stream-disk interaction may lead to thickenings of the disk rim at the
orbital phases $\phi \sim 0.2, 0.5$ and $0.8$, which could appear in eclipse
maps as arcs of enhanced emission due to the reprocessing of radiation from
the inner disk. However, this is not a plausible explanation for the observed
asymmetric arcs because (i) there is no conspicuous bright spot at the disk
rim and no enhanced emission along the gas stream trajectory to signal
any relevant stream-disk interaction, (ii) there is no asymmetric structure
at phase 0.5 and (iii) the asymmetric arcs are well within the disk (at
$R\simeq 0.75\,R_{\rm d}$) and not at its rim.

Alternatively, \citet{Smak01} and \citet{Ogilvie02} proposed
that these asymmetric arcs are caused by irradiation of tidally thickened
sectors of the outer disk by the hot, inner disk regions. However,
eclipse mapping results offer no support for these explanations
\citep{Baptista05}. First, the eclipse mapping procedure projects
any emission produced at height $z$ above the disk back along the
inclined line of sight to the point where it pierces the disk
plane, at a distance $z \tan i$ away from the secondary star with
respect to its true position \citep{Baptista95}. For a high
inclination system such as CTRS~J0934, even a small vertical
height $z\simeq 0.03\,R_{L1}= 1.5\times 10^9$~cm would be enough
to shift both arms by $\Delta x \simeq 0.2\,R_{L1}$ towards the
secondary star with respect to their positions in the eclipse
maps. The resulting highly asymmetric geometrical configuration
would be inconsistent with the framework of tides induced by the
secondary star, which are point-symmetric with respect to the WD
both for the shocks \citep{SS99, Makita00} and the irradiation
\citep{Smak01, Ogilvie02} interpretations. Second, the irradiation
model predicts that irradiation affects mainly the side of the
thickened disk facing the hot inner regions. When the
thickened disk structure is viewed from a highly inclined line
of sight towards the secondary star (i.e., around eclipse) the
irradiated arm 2 would be seen mostly face-on, but arm 1 would
suffer self-occultation (or would be seen at a much lower effective
area). Thus, in the irradiation model one expects that arm 2
appears systematically (and possibly significantly) brighter than
arm 1 around eclipse (and this brightness difference would be
transported to any eclipse map) -- which is not supported by 
eclipse mapping results \citep{Baptista05}.

The remaining
explanation is that the asymmetric arcs are
tidally-induced spiral shocks in the accretion disk. 
Numerical calculations by \citet{Sawada86a,Sawada86b} predicted that the
tidal action of the secondary star leads to accretion disks elongated
in the direction perpendicular to the line joining both stars, with the 
appearance of a two-armed spiral shock wave in its outer regions which
propagates to small radii roughly at the local sound speed, $c_s$
\citep[e.g.,][]{Boffin01}. In an inviscid disk, the opening angle of the
spiral arms (i.e., the angle between the shock surface and the direction
of the orbital motion) is of the order of $\tan\theta_s= c_s/v_\phi$
where $v_\phi$ is the local Keplerian velocity \citep[e.g.,][]{Rozyczka89}.
Observational evidence of such spiral arms were
found in the accretion disk of the dwarf nova IP~Peg during outbursts
\citep{Steeghs97,Harlaftis99,Baptista00,Baptista02}. The comparison
between the velocities (Doppler tomography) and locations (eclipse
mapping) reveal that the spiral pattern extends significantly
inwards of the accretion disk (down to $\simeq 0.2\,R_{L1}$) and that
the observed (Doppler) velocities are up to 30 per cent lower than
the corresponding Keplerian velocities, underscoring the shock nature
of these structures. At the time of the above mentioned studies, the
lack of clear evidence of spiral structures in the accretion disks of
nova-likes and quiescent dwarf novae fostered the idea that, in order
to excite the tidally-induced spiral pattern, the accretion disk needs
to be exceedingly large such as during dwarf nova outbursts.

Subsequent numerical simulations and observations led to a different
picture. \citet{Armitage98} attempted to reproduce the spiral arms in
the accretion disk of IP~Peg. The tidally-induced spiral pattern appears
in both the quiescent and outburst cases. The opening angle of the
spirals increases when the disk viscosity parameter
$\alpha$\footnote{Here we adopt the notation of \citet{Shakura73} and
  write the disk viscosity as $\nu= \alpha c_s H$, where $H$ is the
  disk vertical scaleheight.} 
is increased by a factor of 4, from $\alpha=0.15$ in quiescence to 
$\alpha=0.6$ in outburst, and matches those observed in IP Peg.
\citet{Stehle99} consistently found tidally-induced spiral structures in
accretion disk simulations for both the assumed ``quiescence'' (low
$\alpha=0.01$) and ``outburst'' (high $\alpha=0.3$) cases. Low viscosity
leads to tightly wound spiral arms, while high viscosity results in wide
open spiral arms such as those observed. As a consequence of the increased
viscosity, his ``outburst'' disk becomes $\sim 10$ per cent larger than
in ``quiescence''. \citet{Makita00} and \citet{Sato03} investigated
tidally-induced spiral structures from ``cool'' ($\gamma=1.01$) to ``hot''
($\gamma=5/3$) accretion disks. Spiral patterns are present in all cases,
regardless of outer disk radius. Smaller $\gamma$ values (lower
temperatures) resulted in a more tightly wound spiral pattern, and the
spiral shock waves only disappeared when the tidal force from the
secondary star was artificially cut off. Wide open spirals such as those
observed are obtained when radiative cooling is taken into account.
The summary of the numerical simulations is that tidally-induced spiral
pattern is always present in an accretion disk; its opening angle is large
in hot and/or high-viscosity disks (leading to wide open spiral arms
easier to detect observationally) and small in cool and/or low-viscosity
disks \citep[leading to tightly wound spirals which become
indistinguishable from the underlying accretion disk; e.g.,][]{SS99}.
Observational support for this picture comes from the results of
\citet{Baptista05}, who found that the opening angle of the spiral arms
in IP~Peg progressively decreases while its accretion disk cools during
outburst decline. The further discovery of spiral arms in the accretion
disks of the quiescent dwarf nova V2051~Oph \citep{Rutkowski16,Baptista20}
and of the nova-like systems V347~Pup \citep{Thoroughgood05}, V348~Pup
\citep{Saito16} and EC\,21178 \citep{Khangale20,Ruiz-Carmona20} allows
a crucial test of the idea that an exceedingly large disk is required
in order to excite the spiral pattern.

For eclipsing systems, the width of the eclipse $\Delta\phi_E$ can be
used to estimate the disk radius \citep[e.g., ][]{Baptista20} and to
compare it with the tidal truncation radius $R_T$ \citep{Paczynski77},
\begin{equation}
    \frac{R_T}{R_{L1}}= \frac{0.6}{(1+q)} \frac{a}{R_{L1}}\, ,
    \label{eq:tidal}
\end{equation}
where $R_{L1}/a= 0.5-0.227\log q$ \citep{Plavec64}. For V347~Pup
\citep[$q=0.83\pm 0.05$, $\Delta\phi_E= 0.110\pm 0.005$,][]
{Thoroughgood05} we find $R_d/R_T= 0.86$, for V348~Pup \citep[$0.31\leq q
\leq 0.6$, $\Delta\phi_E= 0.087\pm 0.002$,][]{Saito16} we obtain $R_d/R_T
\simeq 0.7$, and for EC\,21178 \citep[$q=0.48 \pm 0.06$, $\Delta\phi_E=
0.092\pm 0.003$,][]{Khangale20} we find $R_d/R_T\simeq 0.7$ as well.
CRTS~J0934 fits nicely into this group, with an estimated $R_I/R_T=0.75$. 
Furthermore, the quiescent dwarf nova V2051~Oph \citep[$q=0.19\pm 0.03$,
$\Delta\phi_E= 0.082\pm 0.003$,][]{Baptista20} provides an extreme case
with $R_d/R_T=0.56$. The results show that these accretion disks are not
particularly large and do not extend up to the region of strong
gravitational perturbation from the secondary star, indicating that
disk size is not a determining factor for the presence of spiral arms.

While it is not possible to estimate the opening angle of the
spirals directly from the eclipse 
map\footnote{As a consequence of the intrinsic azimuthal smearing effect
 of the eclipse mapping method, spiral arms are smeared in the azimuthal
 direction; their trace in the azimuthal intensity distribution tends to
 be at a constant average radius, which cannot be used to estimate the
 opening angle $\theta_s$ of the spirals \citep{Harlaftis04}.},
their orientation (i.e., binary phase $\phi_s$) and azimuthal extent
$\Delta\phi_s$ are well recovered. There is a clear correlation between
$\theta_s$ and ($\phi_s,\Delta\phi_s$): wide open spirals with large
opening angles are mapped as arcs of small azimuthal extent at an
orientation roughly perpendicular to the line joining both stars; $\phi_s$
and $\Delta\phi_s$ both increase as $\theta_s$ decreases and the spirals
wind up. This correlation led \citet{Baptista05} to devise an indirect
way to estimate $\theta_s$ 
from the orientation of the valleys in the azimuthal intensity
distribution, which rotate clockwise as the opening angle decreases.
They found that the orbital phase of lowest intensity (the valley) in
the azimuthal intensity distribution correlates to the opening angle of
the spirals through the expression,
\begin{equation}
\theta_s(\mathrm{degrees}) = \frac{23.25}{\phi_1 (I_\mathrm{min})} - 29.6 \, ,
\end{equation}
where $\phi_1 (I_\mathrm{min})$ corresponds to the first of the two orbital
phases of intensity valleys in the azimuthal intensity distribution.
We fitted a parabola to the intensities around the first valley of the
azimuthal intensity distributions of Fig.~\ref{spiral}a to find
an average value of $\phi_1 (I_\mathrm{min})= 0.47 \pm
0.04$. This leads to an average opening angle of
$\theta_s= 19\degr \pm 4\degr$. 
This estimate can be visually checked
against the expected trace of tidally-induced spiral arms for opening
angles of $\theta_s= 15\degr, 19\degr$ and $30\degr$ (shown as dot-dashed
red lines) in the right-hand panels of Fig.~\ref{figure2}. Spiral arms
with $\theta_s=30\degr$ ($\theta=15\degr$) underestimate (overestimate) the
orientation and azimuthal extent of the observed asymmetric structures,
while that with $\theta_s=19\degr$ provides a reasonably good match to the
binary phases and azimuthal extent of the observed asymmetric structures.
The $\theta_s$ value obtained for CRTS~J0934 is comparable to that
inferred for the spirals in the quiescent accretion disk of V2051\,Oph
\citep[$\theta_s= 21\degr \pm 3\degr$,][]{Baptista20} and 
5-6\,d after of the onset of the outburst in IP\,Peg 
\citep[$\theta_s= 25\degr \pm 3\degr$,][]{Baptista05}.

The detection of wide open spiral density waves in the nova-like
CRTS~J0934 suggests that its accretion disk has a high mass accretion
rate (and, therefore, high temperatures) and/or a high viscosity. Both
characteristics are typical of novalike systems close to the upper side
of the CV period gap.

\section{Summary}
\label{Summary}

Using long-term photometric observations combined with time-resolved light 
curves of the variable star CRTS~SSS100505~J093416-174421, we provide the 
first detailed characterization of this system. We identify it as an 
eclipsing nova-like cataclysmic variable of the VY~Scl subtype with an 
orbital period of 0.16329188~d (3.919005~h). Tomography of the eclipse 
profile shows us that the accretion disk is completely eclipsed. Residual 
emission at mid-eclipse may be due to matter above or below the orbital 
plane, with only a small contribution from a main sequence
secondary star. We find best fit values of $81.5\degr \pm 1.0\degr$
for the orbital inclination and a mass ratio of $q = 0.45 \pm 0.05$.
This, together with reasonable 
assumptions, allowed us to estimate further system parameters. Even 
being total, the eclipses are rather brief. This is due to an accretion 
disk that is smaller than observed in other nova-like variables. Its outer 
radius is only 21 per cent of the component separation or 36 per cent of
the distance between the WD and the L1 point when observed
in the V-band. However, seen in the near infrared, it increases to 0.54
of the WD-L1 distance. The small V-band disk combined with
the high inclination explains the faint absolute magnitude of $M_g = 7.44$
which is close to the faint end of the absolute magnitude distribution for 
nova-like CVs. The eclipse tomography also reveals the presence of two 
asymmetrical structures in the disk that can be interpreted as
tidally-induced spiral arms.

\section*{Acknowledgements}

This paper is largely based on observations carried out at the Observat\'orio 
do Pico dos Dias (OPD), operated by the Laborat\'orio Nacional de 
Astrof\'{\i}sica (LA/MCTI, Brazil, under OPD observing program 2021A-P-04 
(PI: A. Bruch). Supportive observations from TESS, CRTS, and ZTF were also used.
Data from the TESS space mission were downloaded from the MAST data archive 
at the Space Telescope Science Institute (STScI). Funding for the mission is 
provided by the NASA Explorer Program. STScI is operated by the Association of 
Universities for Research in Astronomy, Inc., under NASA contract NAS 5-26555. 
CRTS is supported by the US NSF under grants AST-0909182 and CNS-0540369. 
The work at Caltech was supported in part by the NASA Fermi grant 
08-FERMI08-0025, and by the Ajax Foundation. The CSS survey is funded by 
the National Aeronautics and Space Administration under grant no. NNG05GF22G 
issued through the Science Mission Directorate Near-Earth Objects 
Observations Programme.
ZTF observations are carried out with the Samuel Oschin 48-inch and the
60-inch Telescope at the Palomar Observatory as part of the Zwicky
Transient Facility project. ZTF is supported by the National Science 
Foundation under Grants No. AST-1440341 and AST-2034437, and a collaboration 
including Caltech, IPAC, the Weizmann Institute for Science, the Oskar Klein 
Center at Stockholm University, the University of Maryland, the University 
of Washington, Deutsches Elektronen-Synchrotron and Humboldt University, 
Los Alamos National Laboratories, the TANGO Consortium of Taiwan, the 
University of Wisconsin at Milwaukee, Trinity College Dublin, Lawrence 
Berkeley National Laboratories, Lawrence Liverpool National Laboratories,
and IN2P3, France. Operations are conducted by COO, IPAC, and UW.
R.L.O. was partially supported by CNPq (PQ-315632/2023-2 and 445047/2024-0) 
and FAPESP (2025/01204-6). C.V.R. thanks the support from the Brazilian
Space Agency (PO~20VB.0009) and CNPq (305991/2024-8). A.S.O. acknowledges
the support from FAPESP (2017/20309-7). I.J.L. thanks FAPESP (grant
\#2024/14358-9 and \#2024/03736-2).

\end{document}